\documentstyle[aps,prb,amssymb,twocolumn,epsf]{revtex}

\newcommand{\h}{\hspace*{0.2ex}{}} 
\newcommand{\hh}{\hspace*{0.1ex}{}}
\newcommand{\kB}{k_{\mbox{\hspace*{-0.08ex}\vipt\rm B}}}
\newcommand{\bmf}[1]{\mbox{\boldmath ${#1}$}}
\newcommand{\bmfi}[1]{\mbox{\scriptsize \boldmath $#1$}}

\begin{document}

\title{A Model for the $\lambda$-Transition of Helium$^\dagger$}

\author{Torsten Fliessbach\cite{email}}

\address{Fachbereich Physik, University of Siegen, 
 57068 Siegen, Germany}

%\date{}

\maketitle

\begin{abstract}
Guided by the analogy to the Bose-Einstein condensation of the ideal
Bose gas (IBG) we propose a new model for the $\lambda$-transition of
liquid helium. Deviating from the IBG our model uses phase ordered and
localized single-particle functions. This means that finite groups of
particles are assumed to be phase-locked. These phase correlations can
be related to the singularity at the transition point and to the
occurrence of the superfluid density.

The model leads to the following results:
\begin{enumerate}
\item A possible explanation of the logarithmic singularity of the
specific heat.
\item A characteristic functional form for the superfluid density which
yields excellent fits to the experimental data.
\item A quantitative prediction of a small but nonzero entropy content
of the superfluid component.
\end{enumerate}
\end{abstract}

%\pacs{PACS number: 67.00} 

% he1.tex

\section{Introduction}
\label{s1}

~In 1938 London\cite{lo}  presented his famous suggestion about the
connection between the $\lambda$-transition in ${}^4$He and the
Bose-Einstein condensation in the ideal Bose gas\cite{zi} (IBG). About
ten years later this point of view was confirmed experimentally by
showing the absence of such a transition in ${}^3$He. Theoretically it
was supported by Feynman\cite{fe} who argued that London's view was
essentially correct. Furthermore, the microscopic IBG provides a means
of understanding\cite{pu} some properties of He~II. The actual
properties of helium are, however, in many respects quite different
from the IBG. In view of this we propose a model which is a
modification of the IBG invented such that relevant equilibrium
properties of the real system are reproduced.

The IBG can be defined by the microscopic states
\begin{equation}
\label{e1}
{\mit\Psi}_{\rm IBG}(\bmf{r}_j,n_k)={\cal S} \prod_{\bmfi{k}} \big[\h
\varphi_{\bmfi{k}} \h \big]^{n_k}\,,
\end{equation}
which are the symmetrized (${\cal S}$) product states of
single-particle functions (s.p.f.)\ $\varphi_{\bmfi{k}}$ of momentum
$\hbar\h\hh\bmf{k}$. The states (\ref{e1}) depend on the coordinates
$\bmf{r}_j$ (where $j=1,\ldots ,N$) of the atoms and on the parameters
$n_k=n_{\bmfi{k}}$.  The temperature dependent expectation values
$\langle n_k\rangle$ of the occupation numbers display the phase
transition of the IBG.

Our model is based on two major assumptions:
\begin{enumerate}
\item It starts with an IBG-related {\em guess}\/ for the form of the
many-body states. As in Eq.\ (\ref{e1}), these states depend on
occupation numbers $n_k$, but beyond Eq.\ (\ref{e1}) they contain
correlations due to the use of localized and phase ordered
single-particle functions.

\item The phase transition is introduced phenomenologically by using the
IBG values for $\langle n_k\rangle$. Due to this feature our model is
closely related to the IBG and will therefore be called {\em almost
ideal Bose gas model}\/, or AIBG.
\end{enumerate}

The basic idea of the AIBG for introducing the relevant correlations is
the following: Normally exponential functions $\varphi_{\bmfi{k}} \propto
\exp (\hh {\rm i}\h \bmf{k}\cdot \bmf{r}_j)$ are used for the s.p.f.\
in Eq.\ (\ref{e1}).  Alternatively one may consider sinus functions
which depend on the component $x_j = \hat{\bmf{x}}\cdot \bmf{r}_j$ of
$\bmf{r}_j$ as follows:
\begin{equation}
\label{e2}
\varphi_{\bmfi{k}} \propto \sin (q\h x_j+\phi_j)\,.
\end{equation}
Here $\hat{\bmf{x}}$ is a unit vector in an arbitrary direction, and 
$q=\hat{\bmf{x}}\cdot \bmf{k}$. For these s.p.f.\  we introduce the
concept of {\em phase ordering}\/ (p.o.):
\begin{equation}
\label{e3}
\phi_j = \phi_0 \qquad \mbox{(phase ordering)}\,.
\end{equation}
Phase ordered (p.o.) s.p.f.\  with the same $q$ (but in general
different $\bmf{k}$) are correlated because their squares
$|\varphi_{\bmfi{k}} ({\bmf{r}}_j)|^2$ exhibit common extrema. Thus
p.o.\ leads to an extra spatial correlation between all atoms with the
same $q$. Such correlations may be important if many atoms share the
same momentum.  Therefore p.o.\ is a potentially decisive correlation
near the transition point in an IBG-like model.

A p.o.\ may be introduced by physical (or Dirichlet) boundary conditions
at the walls of the macroscopic volume $V$. Then the resulting effects
are surface effects and vanish like $V^{-1/3}$ for $V\to\infty$. In
contrast to this, we obtain {\em finite}\/ effects by assuming s.p.f.\ 
$\varphi_{\bmfi{k}}({\rm l.p.o.})$ which are not only phase ordered (p.o.)
but also {\em localized}\/ (l.). For this purpose the volume $V$ is
thought to be divided into $V/V_0$ finite boxes of size $V_0$. The
$\varphi_{\bmfi{k}}({\rm l.p.o})$ are then the s.p.f.\  which are
confined to one of these boxes and subject to Dirichlet boundary
conditions. For these s.p.f.\  we construct a state of the form
(\ref{e1}):
\begin{equation}
\label{e4}
{\mit\Psi}_{\rm M}(\bmf{r}_j,n_k,N_0)={\cal S}\,\prod_{\rm vol}
 \prod_{\bmfi{k}}\,\big[\h \varphi_{\bmfi{k}}({\rm l.p.o.})\h
\big]^{n_k}\,.
\end{equation}
The product runs over all momenta and all finite volumes. The states
${\mit\Psi}_{\rm M}$ depend on the additional parameter $N_0$ which denotes the
number of atoms in $V_0$. For $N_0\to\infty$ these states reduce to IBG
states. As a surface effect of the finite boxes the l.p.o.\ introduces
correlations of the order $N_0^{\,-1/3}$.

As it stands, the states ${\mit\Psi}_{\rm M}$ cannot be used together
with a realistic Hamiltonian $H$. Therefore, we multiply them by a
suitable Jastrow factor,
\begin{equation}
\label{e5}
{\mit\Psi} (\bmf{r}_j,n_k,N_0) = F\h\hh {\mit\Psi}_{\rm M}, \qquad 
F = \prod_{i<j} f(r_{ij}) \,.
\end{equation}
We use Jastrow functions $f(|\bmf{r}_i-\bmf{r}_j|)$ which have been
determined\cite{mc} by minimizing the energy $E_0=\langle F|H|F
\rangle$.

We have now sketched the basic idea of the AIBG, namely the
modification of the IBG due to a phase ordering of the s.p.f.  The
resulting model is rather close to the IBG; in spite of some formal
similarity of the underlying states it is not related to the
quasi-particle model (note that $\sum n_k = N$ for Eq.\ (\ref{e5})). In
Sec.\ \ref{s2} the AIBG will be defined in detail. This includes the
evaluation of the energy
\begin{equation}
\label{e6}
E(n_k) = \big\langle F\h\hh {\mit\Psi}_{\rm M}\hh \big|\hh H \hh \big|
\hh F\h\hh {\mit\Psi}_{\rm M} \big\rangle
\end{equation}
Evaluated with $\langle n_k\rangle_{\rm IBG}$ this energy yields a
logarithmic singularity (Sec.\ \ref{s3}).

Since our approach is based on the IBG (above point 2), the critical
exponent $\beta$ of the model condensate density has the value $1/2$.
This differs from the critical exponent $\nu \simeq 1/3$ of superfluid
density $\rho_{\rm s}$. Therefore, $\rho_{\rm s}$ cannot be identified
with the square of the condensate wave function. In the AIBG we argue
that due to p.o.\ also noncondensed particles contribute to $\rho_{\rm
s}\hh $; in any IBG-like model ($\beta = 1/2$) such a contribution will
be required in order to account for the experimental value $\nu \simeq
1/3$. This contribution leads to the peculiar consequence that the
superfluid component has a nonvanishing entropy $S_{\rm s}\ne 0$. A
prediction for $S_{\rm s}\ne 0$ will either falsify the model right
away, or provide a crucial test of the model. Therefore we extend the
model in Sec.\ \ref{s5} to \ref{s6} such that $\rho_{\rm s}$ and
$S_{\rm s}$ can be calculated.
 
Section \ref{s4} shows how the localization of the s.p.f.\  can be
reconciled with a macroscopic p.o.\ and introduces generalized s.p.f.\
which are able to describe the coherent motion of noncondensed
particles with the condensate. Section \ref{s5} presents the model
expression for $\rho_{\rm s}$ and shows that it yields excellent fits
to the experimental data. With the model parameters fixed by this fit
we obtain in Sec.\ \ref{s6} the crucial model prediction for $S_{\rm
s}\ne 0$.  This prediction is compared to the available experimental
data; it is found that the prediction is at the border of present-day
experimental detectability.

% Section 2

\section{The almost ideal Bose gas model (AIBG)}
\label{s2}

For a complete definition of our model we define the l.p.o.\ s.p.f.\ 
(Sec.\ \ref{s2.1}), calculate the energy $E$ (Sec.\ \ref{s2.2}) and
specify the assumptions about the expectation values of the parameters
(Sec.\ \ref{s2.3}).

\subsection{Localized, phase ordered single-particle functions}
\label{s2.1}

The l.p.o.\ s.p.f.\  in Eq.\ (\ref{e4}) are formally defined as
follows: The macroscopic volume $V$ is divided into $V/V_0$ cubic boxes
of size $V_0$. Within each box we define orthonormalized s.p.f.\  which
obey Dirichlet boundary conditions at the walls:
\begin{equation}
\label{e7}
\varphi_{\bmfi{k}}({\rm l.p.o.}) = \left\{ \begin{array}{ccl}
\displaystyle \sqrt{\frac{8}{V_0}}\, \prod^3_{i=1} 
 \sin (k_i\h x_i)
 & & \mbox{for }\bmf{r}\in V_0\,,  
\\[6mm] 0 && \mbox{elsewhere}\,.
\end{array} \right. \,\,
\end{equation}
The Cartesian components $k_i$ of $\bmf{k}$ are restricted to the
discrete values $q_n$,
\begin{equation}
\label{e8}
q_n = n\cdot \Delta  k,\qquad n=1,2,\ldots , \qquad \Delta k
= \pi/V_0^{\,1/3} \,.
\end{equation}
The many-body states of the AIBG are now defined by Eq.\ (\ref{e5})
with Eqs.\ (\ref{e4}) and (\ref{e7}).

This construction is chosen for simplicity. It contains artificial
aspects like the cubic shape and the identical size of the boxes. The
underlying physical picture is that there are finite regions (of
various shapes and sizes) in which the atoms lower their free energy
(see Sec.\ \ref{s4.1}) by assuming p.o.\ in just {\em one}\/ direction. For
macroscopic physical quantities we have to average (explicitly or
implicitly) over such regions, or over the boxes used for Eq.\ (\ref{e7}).
Formally we perform this averaging by replacing the discrete momentum
sums by integrals:
\begin{equation}
\label{e9}
\sum_{q_n} \;\ldots\quad\Longrightarrow\quad 
\sum_{\, q}\hspace*{-5.6mm}
\int \;\;\ldots \; = \frac{1}{\Delta k} \int_0^\infty \! dq\;\ldots \,.
\end{equation}
The justification of the lower integral bound will be discussed in
Sec.\ \ref{s4.2}. The averaging over boxes of finite size and shape
restores also the translational and rotational invariance of the system
as a whole. The occupation number for $\varphi_{\bmfi{k}}$ will therefore
depend only on $k$ but not on the direction of $\bmf{k}$.

\subsection{Energy}
\label{s2.2}

We evaluate the energy Eq.\ (\ref{e6}) for the Hamiltonian
\begin{equation}
\label{e10}
H=-\sum_i \frac{\hbar^2}{2\h m} \,\Delta_i + \sum_{i<j} u(r_{ij})
\end{equation}
with a realistic atom-atom potential $u(r)$, for example of
Lennard-Jones type.

The ground state (g.s.) may be approximated\cite{mc} by ${\mit\Psi}_0=F$. The
corresponding g.s.\ energy $E_0$ can be written as
\begin{equation}
\label{e11}
E_0(V,N) = \langle F|H|F\rangle 
         =\frac{N^2}{2\h V} \int \! d^3r\;\tilde u(r)\,g_0(r) \,,
\end{equation}
where $g_0(r)$ is the pair correlation function for ${\mit\Psi}_0$ and
$\tilde u(r)$ is given by
\begin{equation}
\label{e12}
\tilde u(r) = u(r) -\frac{\hbar^2}{2\h m}\,\ln f(r)\,.
\end{equation}
Sensible results for $E_0$ and $g_0(r)$ are obtained\cite{mc} by a
simple ansatz for $f(r)$.

The following evaluation of the energy $E$ for ${\mit\Psi}
=F{\mit\Psi}_{\rm M}$ is based on the fact that the main correlations
are determined by $F$. The additional correlations due to
${\mit\Psi}_{\rm M}\ne 1$ are small, they are of the order
$N_0^{\,-1/3}\ll 1$ for the relevant p.o.\ effects. The pair
correlation function $g(r)$ for ${\mit\Psi} = F\h {\mit\Psi}_{\rm M}$
can therefore be written as
\begin{equation}
\label{e13}
g(r,n_k) =g_0(r) +\Delta g(r,n_k),\qquad \Delta g \ll 1\,.
\end{equation}
We relate $\Delta g$ to the pair correlation function $g_{\rm M}$ of
${\mit\Psi}_{\rm M}$. (An explicit expression for $g_{\rm M}$ and details of
the following derivation are given in Ref.\ \onlinecite{f1}). Writing
$g_{\rm M}=1+\Delta g_{\rm M}$ it follows that $\Delta g\to 0$ for
$\Delta g_{\rm M}\to 0$, or $\Delta g\propto \Delta g_{\rm M}$. To
lowest order in $f(r)$ Eq.\ (\ref{e5}) yields
\begin{equation}
\label{e14}
\Delta g(r,n_k)\simeq f(r)^2\,\Delta g_{\rm M}(r,n_k)\,.
\end{equation}
An alternative approximation is $\Delta g\simeq g_0\,\Delta g_{\rm M}$;
the difference to Eq.\ (\ref{e14}) marks the uncertainty in this
approximation. For our purpose both approximations will turn out to be
nearly equivalent. The relation between $\Delta g$ and $\Delta g_{\rm
M}$ cannot be evaluated exactly; for a discussion of this problem see
Ref.\ \onlinecite{bl}.

The potential energy for $F\h {\mit\Psi}_{\rm M}$ is $E_{\rm pot} =
(N^2/V)\,\cdot $ $\int\! d^3r\, g\h\hh u$. The kinetic energy can be
divided into two parts: The Jastrow factors are taken into account by
replacing $u$ in $E_{\rm pot}$ by $\tilde u$, as in Eq.\ (\ref{e11}).
The kinetic contributions due to ${\mit\Psi}_{\rm M}\ne 1$ lead to
$\sum\varepsilon_k\h n_k$ where $\varepsilon_k = \hbar^2 k^2/2\hh m$. Here
$m$ is an effective mass (Ref.\ \onlinecite{f1}) but for simplicity we
do not introduce a new symbol. We emphasize that $m\ne 0$ for $k\to 0$.
This can be related to the fact that the static structure function
$S(k,T)$ is finite for $k\to 0$ and $T\simeq T_\lambda$. The complete
result reads
\begin{eqnarray}
\label{e15}
E(V,N,n_k)&=& 
\big\langle F\h\hh {\mit\Psi}_{\rm M}\hh \big|\hh H \hh \big|
\hh F\h\hh {\mit\Psi}_{\rm M} \big\rangle
\nonumber\\[1.5mm]
&=& \frac{N^2}{2\h V}  \int\! d^3r\; g(r,n_k)\,\tilde u(r) + 
 \sum_{\bmfi{k}} \varepsilon_k\,n_k 
\nonumber\\[1mm]
&=& E_0(V,N) + E_{\rm M}(V,N,n_k)\,.
\end{eqnarray}
The energy $E_{\rm M}=E-E_0$ is obtained by inserting Eqs.\
(\ref{e13}), (\ref{e14}) and (\ref{e11}):
\begin{eqnarray}
\label{e16}
E_{\rm M}(V,N,n_k) &\simeq& \sum_{\bmfi{k}} \varepsilon_k\,n_k 
\\ && {\,} + \, \frac{N^2}{2\h V} \int\! d^3r\; f(r)^2 \,\Delta g_{\rm
M}(r,n_k)\,\tilde u(r)\nonumber\,.
\end{eqnarray}
The decisive contributions in $\Delta g_{\rm M}$ are long-ranged
because the low momenta dominate near the transition. For the intended
evaluation of the critical behaviour we may therefore use the
approximation
\begin{equation}
\label{e17}
\Delta g_{\rm M}(r,n_k)\simeq \Delta g_{\rm M}(0,n_k)\,.
\end{equation}
This yields
\begin{equation}
\label{e18}
E_{\rm M}(V,N,n_k) =\sum_{\bmfi{k}}
 \varepsilon_k\,n_k + \frac{N}{2}\,\frac{w_0}{v}
\;\Delta g_{\rm M}(0,n_k)\,,
\end{equation}
where $v=V/N$ and
\begin{eqnarray}
\label{e19}
\frac{w_0}{v} &\simeq &
\frac{1}{v} \int\! d^3r \,f(r)^2\,\tilde u(r) \simeq
\frac{1}{v} \int\! d^3r \,g_0(r)\,\tilde u(r) 
\nonumber\\[2mm]
&\simeq &
\frac{2\h E_{\rm g.s.}}{N}= -14.3\, \kB  \hh {\rm K} \,.
\end{eqnarray}
Using McMillan's parameters\cite{mc}, the first integral yields
$-14.8\,\kB\hh {\rm K}$. Within the uncertainty of Eq.\ (\ref{e14})
it may be replaced by the second integral. This integral is the
theoretical g.s.\ energy (\ref{e11}). By equating this energy to the
experimental ground state energy $E_{\rm g.s.}$ we fix the value of
$w_0/v$.

Up to a constant, the energy $E_{\rm M}$ (\ref{e18}) can be expressed by
\begin{eqnarray}
E_{\rm M} &=& 
\big\langle \hh {\mit\Psi}_{\rm M}\hh \big|\hh H \hh \big|
\hh  {\mit\Psi}_{\rm M}\hh \big\rangle
 \quad \mbox{with} 
\nonumber\\[1mm]
\label{e20}
H_{\rm M} &=& -\sum_i\frac{\hbar^2}{2\h m}\,\Delta_i + 
w_0\sum_{i<j}\delta(\bmf{r}_i-\bmf{r}_j)\,.
\end{eqnarray}
This means that within sensible approximations the energy $E_{\rm M}$
in $E=E_0+E_{\rm M}$ can be calculated as the expectation value of a
simple product state ${\mit\Psi}_{\rm M}$ with a simple Hamiltonian $H_{\rm
M}$. It should be noted that $H_{\rm M}$ is solely introduced for
evaluating $E_{\rm M}$, it cannot serve as a model Hamiltonian in other
respects. In particular the stability of the system is guaranteed by
the part $E_0$ of the total energy $E=E_0+E_{\rm M}$; the corresponding
part of the Hamiltonian is not contained in $H_{\rm M}$.

The attractive zero-range force in $H_{\rm M}$ has the following
meaning: The l.p.o.\ implies an enhanced probability of finding two
atoms near together. Due to the Jastrow factors in Eq.\ (\ref{e5}) this
enhancement is effective only in the range where the realistic
interaction is attractive; the strength of this attraction is measured
by $w_0/v$. A zero-range force can be used because for $g_{\rm M}$ the
enhancement is approximately $r$-independent up to a few {\AA}.

The evaluation of Eq.\ (\ref{e20}) is straightforward. The two-body
matrix elements $w_{\bmfi{k\hh q}}$ of the $\delta$-force with the
s.p.f.\ (\ref{e7}) are given by
\begin{eqnarray}
\label{e21}
w_{\bmfi{k\hh q}} &=&
\big\langle \h \varphi_{\bmfi{k}}\h \varphi_{\bmfi{q}}\h \big| \h w_0\,
\delta(\bmf{r}_1-\bmf{r}_2)\h \big|\h \varphi_{\bmfi{k}}\h
\varphi_{\bmfi{q}} \big\rangle 
\nonumber\\[1.5mm]
&=&
\frac{w_0}{V_0}\,\,\prod_{i=1}^3\left[
\,1+\frac{1}{2}\,\delta_{k_iq_i}\,\right] 
\nonumber \\ &=&
\frac{w_0}{V_0} \bigg[ 1+\frac{1}{2} \,\sum_{i=1}^3\delta_{k_iq_i}
+\ldots\,\bigg] \,.
\end{eqnarray}
The somewhat artificial construction Eq.\ (\ref{e7}) implies p.o.\ in 3
orthogonal directions. In contrast to this we assume that the p.o.\ is
actually realized in just one direction. Therefore we omit products of
$\delta$-functions in the last expression of Eq.\ (\ref{e21}) which
originate from simultaneous p.o.\ in 2 or 3 directions. The remaining
$\delta$-terms describe the mutual correlation of all particle with the
same momentum $q$ in one direction. The number $\nu_q$ of these
particles is
\begin{equation}
\label{e22}
\nu_q =\sum_{k_2,k_3} n_k \quad\mbox{where~~ } 
k=|\bmf{k}| = \sqrt{ q^2 + k_2^{\,2} +k_3^{\,2}}\,.
\end{equation}
The obvious form of the correlation energy is $E_{\rm corr}\propto \sum
\nu_q^{\,2}$. The actual result follows from Eq.\ (\ref{e20}) with Eqs.\
(\ref{e21}) and (\ref{e22}):
\begin{equation}
\label{e23}
E_{\rm corr}(V,N,n_k)= \frac{3\,w_0}{2\h v} \frac{1}{N_0^{\,1/3}}
\bigg[\frac{1} { N^{2/3}} \Big( \sum_{q} \hspace*{-5.6mm} \int
\;\,\nu_q^{\,2}+2\h\hh n_0\h \nu_0 \Big)\bigg]\h .
\end{equation}
For the replacement (\ref{e9}) of the sum by an integral we took into
account a possible finite condensate fraction $n_0/N$. The index zero
of $n_0$ and $\nu_0$ stands for the lowest possible momentum value. 

The complete expression for the energy $E$ of Eq.\ (\ref{e15}) becomes
\begin{equation}
\label{e24}
E(V,N,n_k) =E_0 + E_{\rm M} = E_0 + E_{\rm IBG} +E_{\rm corr}\,.
\end{equation}
Here $E_0=E_0(V,N)$ is given by Eq.\ (\ref{e11}) and $E_{\rm IBG}$ by
\begin{equation}
\label{e25}
E_{\rm IBG}(V,N,n_k) =\sum_{\bmfi{k}} \varepsilon_k\,n_k\,.
\end{equation}
The critical contribution in Eq.\ (\ref{e24}) is the correlation
energy $E_{\rm corr}$ (\ref{e23}). In the evaluation of Eq.\
(\ref{e20}) leading to Eq.\ (\ref{e24}) we dismissed a constant term
stemming from the constant in Eq.\ (\ref{e21}), the higher order terms
($\propto N_0^{-2/3}$ or $N_0^{-1}$) because they come from p.o.\ in
more than one direction, and a term $\propto n_0^{\,2}$ because it is
noncritical (that means eventually negligible near the transition
point).

For the finite boxes the momentum sums have to be evaluated with
$\Delta k = \pi /V_0^{\,1/3}$; the subsequent summation over all boxes
yields, however, a factor $\h V/V_0$. For plain momentum sums (without
$\delta_{k_iq_i}$-terms) this procedure can be abbreviated by
performing the common summation with $\Delta k = \pi/V^{1/3}$. The
final momentum sums ins Eqs.\ (\ref{e23}) and (\ref{e25}) are performed
in this common way.  This implies that the occupation numbers $n_k$ do
no longer refer to a finite box (as in Eq.\ (\ref{e4})) but to the
macroscopic system (with $V$ and $N$). In Eq.\ (\ref{e23}) this implies
$\sum\nu_q^{\,2} \propto N^{5/3}$ and an $N$-independent $E_{\rm
corr}/N$.

\subsection{Statistical assumptions}
\label{s2.3}

~~In order to calculate the thermodynamic energy $E(T, V, N)$ from
Eq.\ (\ref{e24}) we need the temperature dependent expectation values
of the parameters of ${\mit\Psi}$. Our model uses the IBG expression for
$\langle n_k\rangle$ and treats $N_0$ as an adjustable constant. We
discuss and specify these assumptions.

Any model of the $\lambda$-transition except the IBG introduces the
phase transition phenomenologically. In the AIBG this phenomenological
assumption is the use of the expectation values $\langle
n_k\rangle_{\rm IBG}$. This assumption can be made plausible (to some
extent) by observing that the additional contribution $E_{\rm corr}$ in
$E_{\rm M}=E_{\rm IBG} +E_{\rm corr}$ is small in the sense
\begin{equation}
\label{e26}
\frac{E_{\rm corr}}{E_{\rm IBG}} = {\cal O}(y)\quad \mbox{where} \quad
y = \frac{|w_0/v|\, N_0^{\,-1/3}} {\kB T_\lambda}\ll 1\,.
\end{equation}
The actual parameter values will yield $y=0.13$, Eq.\ (\ref{e42}). For
$N_0\to\infty$ or $y\to 0$ one obtains $E_{\rm M}\to E_{\rm IBG}$ and
thus $\langle n_k\rangle_{\rm IBG}$. (The $n_k$-independent
contribution $E_0(V,N)$ in $E$ is without influence on the expectation
values).  Therefore, we expect $\langle n_k\rangle =\langle
n_k\rangle_{\rm IBG} \,[\,1 +{\cal O}(y)\,]$. The evaluation of $E_{\rm
corr}$ with $\langle n_k \rangle_{\rm IBG}$ is then valid up to first
order in $y$; it is this contribution which yields the logarithmic
singularity. Another argument for using the $\langle n_k\rangle_{\rm
IBG}$ is that the additional term $\sum \nu_q^{\,2}$ correlates many
s.p.\  states with many others, and that it will be without much
influence on the occupation of a particular s.p.\  state.

The AIBG assumes expectation values $\langle n_k\rangle$ of the
IBG-form,
\begin{eqnarray}
\label{e27}
\langle n_k\rangle &=&
\frac{1}{\exp \,[(\,\varepsilon_k-\mu )/\kB T] -1}
\nonumber\\[1mm] 
&=&\frac{1}{\,\exp \,(x^2+\tau^2)-1}\,.
\end{eqnarray}
Here $\mu$ is the chemical potential and $\kB $ is Boltzmann's constant.
We have introduced the dimensionless quantities $\tau^2=-\mu /\kB T$ and
\begin{equation}
\label{e28}
x = \frac{\lambda\h\hh |\bmf{k}|}{\sqrt{4\hh \pi }}\quad\mbox{with}\quad
\lambda =\frac{2\hh \pi\hbar}{\sqrt{2\hh \pi\h m \h \kB T}}\,.
\end{equation}
The transition temperature of the IBG is given by the following
condition for the thermal wave length $\lambda =\lambda (T)$:
\begin{equation}
\label{e29}
\lambda (T_\lambda ) =\left[ \,v \,\zeta (3/2)\,\right]^{1/3}\,,
\end{equation}
where $\zeta (3/2) =2.6124$ denotes Riemann's zeta function. In
applying the AIBG to the real system we will identify $T_\lambda$ with
the actual transition temperature; formally this can be achieved by
inserting a suitable effective mass $m=m(v)$ in $\lambda$ of Eq.\
(\ref{e28}), Ref.\ \onlinecite{fe}. In the following we use the
relative temperature
\begin{equation}
\label{e30}
t=\frac{T-T_\lambda }{T_\lambda }\,.
\end{equation}
For $t\ge 0$ the chemical potential is determined by the particle
number condition $\Sigma\hspace*{-2.2mm}\raisebox{.25mm}{$\int$}\h
n_k = N$. For $t\to 0$ this condition yields $\mu -\varepsilon_0 \to 0$
and thus $n_0\to\infty$. For simplicity we rename $\mu -\varepsilon_0$ by
$\mu$; it is this new $\mu$ which vanishes at $t=0$ also for
$\varepsilon_0\ne 0$.

We are interested in the critical properties. Therefore we expand $\mu$
or, equivalently, $\tau$ for $|t|\ll 1$:
\begin{equation}
\label{e31}
\tau (t) =\sqrt{\frac{-\mu}{\kB T}}=\left\{ \begin{array}{ccc}
a\,t+b\,t^2 +\ldots &\quad& (t>0)\,, \\[1.2mm]
 a'|t| +b\/'t^2+\ldots &&(t<0)\,. \end{array} \right.
\end{equation}
This expansion is directly related to the model condensate fraction
\begin{eqnarray}
\label{e32}
\frac{\langle n_0\rangle}{N} &=& 1-\frac{1}{N} \sum\hspace*{-5.3mm}
\int \,\,\,\langle n_k \rangle 
\nonumber \\[1.5mm] &=&
\left\{ \begin{array}{ccc}
 0 && (t>0), 
\\[1.2mm] 
f\,|t| +g\,t^2+\ldots &&(t<0).
\end{array} \right.
\end{eqnarray}
The coefficients $f,\,g,\ldots$ are determined by $a',\, b{}',\ldots$;
in particular $f=3/2 +2\sqrt{\pi }\,a'/\zeta (3/2)$. The IBG yields
finite values for $a,\;b,\ldots\;$, and $a'=b\/'=\ldots =0$.

Deviating from the IBG, the AIBG admits a coefficient $a'\ne 0$ in Eq.\
(\ref{e31}) or $f>3/2$ in Eq.\ (\ref{e32}). This deviation is
introduced phenomenologically for the following reason: For $t>0$ the
evaluation of $\sum \nu_q^{\,2}$ with $\langle n_k \rangle_{\rm IBG}$
yields straightforwardly a logarithmic singularity. This singularity
can be directly continued to $t<0$ by admitting $a'\ne 0$; the
singularity itself does not depend on the value for $a'$. The form of
the expansion (\ref{e31}) is to some extent plausible because of its
simplicity and symmetry.  Physically $a'\ne 0$ means an energy gap of
$a'\/^2t^2\kB T$ for the condensed s.p.\  state for $t<0$.

For evaluating the thermodynamic energy we need also the expectation
values of products of $n_k$'s. (We do not need, however, the full
statistical information contained in the density matrix). For such
products the IBG yields
\begin{equation}
\label{e33}
\langle n_{\bmfi{k}}\h n_{\bmfi{q}}\rangle =\langle n_{\bmfi{k}} \rangle
\langle n_{\bmfi{q}}\rangle \qquad (\bmf{k}\ne \bmf{q})\,.
\end{equation}

Compared to the IBG, the AIBG states depend on one additional parameter,
namely $N_0$. In principle, this parameter (as well as the occupation
numbers) should be determined from the condition of minimal free energy
$F$ yielding a temperature dependent expectation value $\langle
N_0\rangle$. Instead of this, we treat $N_0$ as an adjustable constant.
Simplified estimates (Sec.\ \ref{s4.1} and Ref.\ \onlinecite{f1}) yield
a sensible value for $N_0$ at $t=0$.

% section 3

\section{Logarithmic singularity}
\label{s3}

We evaluate the thermodynamic energy $E(T,V,N)=\langle E(V,N,n_k)
\rangle$. According to Eq.\ (\ref{e24}) it is of the form
\begin{equation}
\label{e34}
E(T,V,N) = E_0(V,\!\h N) + E_{\rm IBG}(T,\!\h V,\!\h N) + E_{\rm
corr}(T,\!\h V,\!\h N) \hh .
\end{equation}
The term $E_0(V,N)$ is essential for a realistic compressibility but it
does not contribute to the specific heat. The energy $E_{\rm IBG}=\sum
\varepsilon_k\h \langle n_k\rangle$ is calculated as in the IBG. Equations
(\ref{e23}) and (\ref{e33}) yield for the decisive contribution $E_{\rm
corr}$,
\begin{equation}
\label{e35}
\frac{E_{\rm corr}(T,\!\h V,\!\h N)}{N} = \frac{3\hh w_0}{2\h v} 
\frac{1}{N_0^{\,1/3}} \frac{1}{N_{}^{5/3}} 
\bigg[ \!\sum\hspace*{-5.3mm}\int \,\h\langle 
\nu_q\rangle^2 + 2\h \langle n_0\rangle \langle\nu_0\rangle
\!\h\bigg]\!\h .
\end{equation}
Leaving away unessential constants we sketch how $E_{\rm corr}$ yields a
logarithmic singularity: For $|t|\to 0$ and $x\to 0$ the occupation
numbers (\ref{e27}) behave like $\langle n_x\rangle \sim 1/[x^2+t^2]$ and
$\langle \nu_q\rangle \sim \int \!dk\,k\,(k^2+q^2+t^2)^{-1}\sim \ln
(q^2+t^2)$. It follows that $\langle n_0\rangle\langle\nu_0\rangle \sim
|t|\ln |t|$ and $\Sigma \hspace*{-2.2mm}\raisebox{.3mm}{$\int$}\langle
\nu_q\rangle^2 \sim \int \!dq\,[\,\ln (q^2+t^2)\,]^2 \sim \mbox{const.}
+ |t| \ln |t|$.

We present now the results of the detailed calculation. The $\langle
\nu_q\rangle$ of Eq.\ (\ref{e22}) with Eq.\ (\ref{e27}) can be
evaluated analytically,
\begin{eqnarray}
\label{e36}
\frac{\langle \nu_q\rangle}{N^{2/3}}
&=& \frac{1}{N^{2/3}} \!\!\!
\sum_{\,\,\,\,\,\,\,\,k_2,k_3}\hspace*{-8.5mm} 
\int\,\,\,\,\,\langle n_k\rangle 
\nonumber\\[1mm] &=&
-\,\frac{v^{2/3}}{\lambda^2} \, \ln \left[ \h 1-\exp
\left(-x^2-\tau^2\right)\right]\,.
\end{eqnarray}
We used the dimensionless quantities $\tau$ of Eq.\ (\ref{e31}) and
$x = q\hh \lambda/\sqrt{4\pi}$. The integral in Eq.\ (\ref{e35}) becomes
\begin{equation}
\label{e37}
\frac{1}{N^{5/3}}\, \sum\hspace*{-5.3mm}\int\,\,\,\langle\nu_q\rangle^2
=\frac{2}{\sqrt{\pi}}\; \zeta(3/2)^{-5/3} \left(
\frac{T}{T_\lambda} \right)^{\!\!\h 5/2} L(\tau )\,,
\end{equation}
where $L(0)=8.30$ and
\begin{eqnarray}
\label{e38}
L(\tau ) &=& \int_0^\infty\! dx\, \Big( 
\ln \left[ \h 1-\exp \left(-x^2-\tau^2\right)\right]
\Big)^{\! 2}
\\[1.5mm]
&=& L(0) + 4\hh \pi\h\tau\hh \ln \tau + 
4\hh \pi\h\tau\h \big[ \ln (2) - 1 \h\big] + {\cal O}(\tau^2) \,. 
\nonumber
\end{eqnarray}
The expectation value $\langle \nu_0\rangle$ is given by Eq.\
(\ref{e36}) with $x=0$ because $\varepsilon_0$ has been absorbed in
$\tau^2=-(\mu -\varepsilon_0)/\kB T$ (remark after Eq.\ (\ref{e30})).
Using $\langle \nu_0\rangle \sim \ln |t|$, $\langle n_0\rangle \sim
|t|$ and $\tau \sim |t|$ we see that both terms in Eq.\ (\ref{e35})
yield a contribution $\sim t\cdot \ln|t|$. The critical ($|t|\ll 1$)
behaviour of the specific heat reads therefore
\begin{eqnarray}
\label{e39}
c_V(T,V) &=& \left( \frac{\partial (E/N)}{\partial T}\right)_{\!V} 
\\[1.5mm]
 &=& \left\{ \begin{array}{lcc} -A\cdot \ln |t| + B +\ldots && (t>0)\,,
\\[1.5mm]
 -A'\cdot \ln |t| + B' +\ldots && (t<0) \,.\end{array}
\right. \nonumber
\end{eqnarray}
The detailed evaluation of Eq.\ (\ref{e35}) yields
\begin{equation}
\label{e40}
A = A' = \h 9\,\hh\zeta (3/2)^{-2/3} \,\frac{|w_0|}{v}\,
\frac{1}{N_0^{\,1/3}}\, \frac{1}{T_\lambda}\,.
\end{equation}
The terms depending on the parameter $a'$ cancel. We insert $|w_0|/v$
of Eq.\ (\ref{e18}) and $T_\lambda =2.17\,{\rm K}$ in the result
(\ref{e40}).

Experimentally a logarithmic singularity has been measured for $c_P$
over several decades by Ahlers\cite{a71}. For saturated vapour pressure
we may use $c_V\approx c_P$. (The consistency of a logarithmic form for
both, $c_V$ and $c_P$, has been demonstrated by Lee and
Puff{\h}\cite{le}).  The comparison of our result (\ref{e40}) with the
experimental values $A\approx A'\approx 0.63\,\kB $ (Eq.\ (48) of Ref.\
\onlinecite{a71}) determines the model parameter $N_0$,
\begin{equation}
\label{e41}
N_0^{\,1/3} \approx 50\qquad\mbox{(adjusted)}\,.
\end{equation}
This fixes also the small parameter $y$ of Eq.\ (\ref{e26}) which is a
quantitative measure for the deviations of the AIBG from the IBG,
\begin{equation}
\label{e42}
y=\frac{|\hh w_0/v \hh |\, N_0^{\,-1/3}} {\kB \hh T_\lambda}\h
\approx 0.13\,.
\end{equation}
Experimentally one finds a small difference between $A$ and $A'$
(Ahlers\cite{a71} reports $5\%$) and an indication of a cut of the
logarithmic singularity (slightly negative critical exponents\cite{a71}
for $c_P$). In the AIBG a difference between $A$ and $A'$ could be
obtained if corrections to approximation (\ref{e17}) are taken into
account which differ for $t>0$ and $t<0$. A possible cut of the
singularity is connected with the lowest possible values of the
momenta.

The coefficients $B$ and $B'$ in Eq.\ (\ref{e39}) can be calculated
from Eq.\ (\ref{e35}), too.  They depend on the value of the parameter
$a'$ which will be determined only later (Sec.\ \ref{s5.3}). Moreover a
generalization of the s.p.f.\  to be introduced in Sec.\ \ref{s4.3} will
yield another contribution to the jump in $B'-B$.

% section 4

\section{Phase ordering}
\label{s4}

We review various aspects of the p.o.\ assumed in the AIBG. First we
show that this p.o.\ is favoured by the free energy (Sec.\ \ref{s4.1}).
Then we review the necessity of the localization of the s.p.f.\  for
obtaining finite correlation effects due to p.o.\ (Sec.\ \ref{s4.2});
this leads to the notion that the coherence range of p.o.\ approaches
infinity at the transition point. Finally (Sec.\ \ref{s4.3}) we connect
the p.o.\ with a potential superfluid flow.

\subsection{Free energy due to phase ordering}
\label{s4.1}

The l.p.o.\ leads to a lower energy, $E_{\rm corr}/N \sim (w_0/v)\cdot$
$N_0^{\,-1/3}<0$. At the same time such an ordering reduces the
entropy, $\Delta S({\rm p.o.})<0$. We show that $\Delta S({\rm p.o.})$
is relatively small and that p.o.\ is therefore favoured by the free
energy.

Arbitrary phases in Eq.\ (\ref{e2}) are statistically equivalent to an
equal weight of the two independent choices for the phase, $\phi_0$ as
in Eq.\ (\ref{e3}) or $\phi_0+\pi /2$. We consider p.o.\ in just one
direction. For each $\bmf{k}$ we have then two s.p.\  states. The
number of possible distributions of $n_k$ bosons on 2 states is
$\,n_k+1$. The p.o.\ requires that all $n_k$ bosons go into the same
s.p.\  state; compared to a statistical distribution this implies the
entropy change $\Delta s({\rm p.o.})=-\kB \ln (n_k+1)$. On the other
hand, the $n_k$ atoms gain by p.o.\ the energy $\Delta e({\rm
p.o.})\sim (E_{\rm corr}/N)\,n_k$.  The free energy change $\Delta f$
is then
\begin{equation}
\label{e43}
\Delta f({\rm p.o.}) \sim -\,\frac{|w_0|}{v\,N_0^{\,-1/3}}\, n_k +
\kB\hh T\h\ln (n_k+1)\,.
\end{equation}
In spite of the smallness (\ref{e42}) of the considered correlations
the p.o.\ is favoured ($\Delta f({\rm p.o.}) < 0$) as soon as $n_k\gg
1$.  The boson property of the atoms is decisive for this conclusion:
It implies that $|\Delta s|$ is not proportional to $n_k$ but only to
$\h \ln n_k\,$; this behaviour may be paraphrased by `bosons like to go
into the same state'. For the relevant lowest s.p.\  states ($k\simeq 0$)
the condition $n_k\gg 1$ is fulfilled for $t\ll 1\h$. Thus p.o.\ will
indeed be adopted when the transition point is approached ($t\to 0^+$).

As a simplification our ansatz (\ref{e4}) assumes p.o.\ for {\em all}\/
s.p.f. Summing over all momenta and including numerical factors one
finds then $E_{\rm corr}\simeq T\Delta S({\rm p.o.})$ for $T\simeq
T_\lambda$ and for the $N_0$ of Eq.\ (\ref{e41}). Reversely, the
condition $E_{\rm corr}\sim T_\lambda\,\Delta S({\rm p.o.})$
constitutes a rough {\em theoretical}\/ estimate for the model
parameter $N_0$, leading to an $N_0$ of the order (\ref{e41}).

\subsection{Phase coherence volume}
\label{s4.2}

In the Introduction we argued that for finite correlation effects due
to p.o.\ the s.p.f.\  have to be localized. We review this point and find
that the localization is not required for the lowest s.p.\  state. We
present then an estimate for the extension of this lowest s.p.f.; within
this extension the p.o.\ will be coherent.

The {\em finiteness}\/ of the correlation energy ($\propto
N_0^{\,-1/3}$) is due to the {\em finite}\/ spacing $\Delta k=\pi
/V_0^{\,1/3}$ of the $q$-values in the sum $E_{\rm corr}\sim \sum
\nu_q^{\,2}$. Nonlocalized s.p.f.\  would lead to the same $E_{\rm
corr}$ provided that only s.p.\  states with
\begin{equation}
\label{e44}
q_n = q_0 + n\cdot \Delta k,\qquad (n=0,1,2,\ldots)\,.
\end{equation}
are occupied. In the macroscopic system the possible $q$-values are,
however, dense, and the entropy drives the particles into the
occupation of all available states. This is the reason why a finite
$\Delta k$ can be realized only for s.p.f.\  localized within a volume
$V_0=(\pi /\Delta k)^3$.

This argument shows that we may either start from finite volumes or
from finite $\Delta k$, the second feature follows from the first one.
There is, however, one difference: Starting from the finite spacing
$\Delta k$ in Eq.\ (\ref{e44}), the natural condition for the lowest
$q$-value is $q_0<\Delta k$. The corresponding s.p.f.\  can then not be
localized within $V_0$; as just stated there is also no such need for a
localization for the required correlation effect.

In view of this we introduce the following modification of the states
(\ref{e4}): Only the s.p.f.\  with $q_{n>0}$ are localized within
$V_0$, the s.p.f.\  with $q_0$ have a larger extension. We present a
crude estimate for the volume $V_{\rm c}$ of the lowest s.p.f.\
$\varphi_0$ which becomes the condensate state for $t<0$: Let $V_{\rm
c}$ be some multiple of $V_0$, that means $V_{\rm c}=WV_0$. In a volume
$V_{\rm c}$ there are $W$ s.p.\  states below $\Delta k$ out of which
only one is occupied. A redistribution of the $n_0$ atoms over these
$W$ states would increase the entropy by $\Delta s \simeq \kB \ln
(n_0)^W$. At the same time these atoms would loose their correlation
energy, $\Delta e \sim n_0 \hh |w_0/v|\hh N_0^{\,-1/3}$. The stability
condition $T_\lambda\,\Delta s <\Delta e$ yields an upper bound for
$W$. Using $\langle n_0(t>0)\rangle \sim t^{-2}$ and $\langle
n_0(t>0)\rangle ={\cal O}(N)$ we obtain
\begin{equation}
\label{e45}
V_{\rm c}(t) = W\,V_0 \sim V_0\cdot \left\{ \begin{array}{ccc}
t^{-2} &\qquad& (t>0)\,, \\[1mm]
\infty && (t<0)\,.
\end{array} \right.
\end{equation}
This result means the following: Without changing our previous
calculations we can modify the states (\ref{e4}) in such a way that the
lowest s.p.f.\  $\varphi_0$ has the large and eventually infinite volume
$V_{\rm c}$, whereas the s.p.f.\  with $q_n>0$ are localized. This
modification of Eq.\ (\ref{e4}) has important consequences:
\begin{enumerate}
\item Within the volume $V_{\rm c}$ the direction of p.o.\ is
defined by $\varphi_0$. Therefore, $V_{\rm c}(t)$ of Eq.\ (\ref{e45})
is the {\em phase coherence volume}\/ of the considered p.o.

\item The p.o.\ constitutes a symmetry breaking because in an infinite
system the atoms are free to adopt arbitrary phases $\phi_j$ in Eq.\
(\ref{e2}). (The average over the size and shape of finite boxes
(assumed in Sec.\ \ref{s2.1}) does not restore this symmetry). The
result (\ref{e45}) means that this symmetry breaking changes its
character from local to global at the transition point. Approaching the
transition point ($t\to 0^+$) the directions of p.o.\ of neighbouring
$V_0$'s get aligned, and for $t<0$ the coherence is potentially
infinite.

\item Strictly finite and separated boxes imply a momentum cut at
$\Delta k$ and consequently a cut of the logarithmic singularity.
This cut would be described by a lower bound $\Delta k$ in the
integral in Eq.\ (\ref{e9}). For Eq.\ (\ref{e44}) we revise the
replacement Eq.\ (\ref{e9}) of the momentum sums by integrals:
\begin{equation}
\label{e46}
\left\langle \;\sum_{q_n=q_0+n\cdot\Delta k} \;\ldots\,\;\right
\rangle_{\!\!\!q_0} \;\simeq \;
\frac{1}{\Delta k} \int_{q_{0,{\rm min}}}^\infty dq \; \ldots\,.
\end{equation}
The brackets indicate an average over possible $q_0$'s with
$q_0\le\Delta k$. The lowest value for $q_0$ is determined by Eq.\
(\ref{e45}). A lower bound $q_{0,{\rm min}}\sim t^2$ (for $t>0$) of the
integral does not lead to a cut of the logarithmic singularity.
\end{enumerate}

\subsection{Complex phase ordering}
\label{s4.3}

The phenomenon of superfluidity can be explained by assuming a
macroscopic wave function $\psi$. The phase ${\mit\Phi} (\bmf{r})$
of the complex $\psi$ yields the velocity $v_{\rm s}=(\hbar/m)\bmf{\nabla}
{\mit\Phi}$ of a potential superfluid flow. In the AIBG the condensed
particles for a macroscopic wave function for $t<0$, too, Eq.\
(\ref{e45}). In order to join the common description we replace the
real s.p.f.\ $\varphi_0$ by the complex s.p.f.:
\begin{equation}
\label{e47}
\bar \varphi_0 = \varphi_0 \,\exp \,[\,{\rm i}\hh\h 
{\mit\Phi}(\bmf{r})]\,.
\end{equation}
This replacement does not affect the previous calculations; moreover,
the Jastrow factors in Eq.\ (\ref{e5}) are without influence on a flow
due to $\bmf{\nabla}{\mit\Phi} \ne 0$. With Eq.\ (\ref{e47}) all results obtained
from the assumption of a macroscopic wave function apply to our model,
too. Besides this basic conformity the AIBG leads, however, also to a
peculiar modification which will be discussed in the following.

In an IBG-like model (with $\langle n_0 \rangle \sim |t|$ or $\beta
=1/2$) we have to assume that noncondensed particles contribute to the
superfluid density $\rho_{\rm s}$ in order to get agreement with the
experiment ($\rho_{\rm s}\sim |t|^{2/3}$ or $\nu =1/3$). The considered
p.o.\ should in some way imply that noncondensed particles move
coherently with the condensate. We introduce the possibility of a net
current of the noncondensed particles by replacing the original real
s.p.f.\  $\varphi_{\bmfi{k}}$ by
\begin{equation}
\label{e48}
\bar \varphi_{\bmfi{k}} = \varphi_{\bmfi{k}}\, 
\exp \left[ \,{\rm i}\hh\h {\mit\Phi}_{\bmfi{k}} (\bmf{r})\right]\,.
\end{equation}
A coherent motion with the condensate can now be described by the
condition ${\mit\Phi}_{\bmfi{k}}={\mit\Phi}$. We call this condition
{\em complex phase ordering}\/ (c.p.o.) named after the complex phase
factors in Eqs.\ (\ref{e47}) and (\ref{e48}). In contrast to this the
p.o.\ (\ref{e3}) will be called real phase ordering (r.p.o.). We note
two points: The spatial correlations of the s.p.f.\  are unchanged by
the replacement (\ref{e47}), (\ref{e48}) because $|\varphi|^2 =
|\bar\varphi|^2$.  Secondly, for a coherent macroscopic flow it is
sufficient that the field ${\mit\Phi}(\bmf{r})$ is macroscopic; it is
then not necessary that {\em all}\/ contributing s.p.f.\ are of
macroscopic range.

The replacement (\ref{e47}) is suggested by the familiar picture of a
superfluid, and Eq.\ (\ref{e48}) with
${\mit\Phi}_{\bmfi{k}}={\mit\Phi}$ is necessary to reconcile $\beta
=1/2$ (model) with $\nu =1/3$ (experiment). The fairly straightforward
introduction of the phase factors (\ref{e47}) and (\ref{e47})
determines the crucial model prediction for $S_{\rm s}\ne 0$. The fit
to the experimental $\rho_{\rm s}$ determines the extent of the c.p.o.\
and consequently $S_{\rm s}$.

Presenting a peculiar prediction like $S_{\rm s}\ne 0$ we feel obliged
to make it {\em quantitative}\/. For this purpose we have to write down
explicit expressions for $\rho_{\rm s}$ and $S_{\rm s}$. This requires
additional assumptions which are introduced by plausibility arguments.
As explained in the last section, the quantitative result for $S_{\rm
s}$ will not depend sensitively on these assumptions because the model
expression for $\rho_{\rm s}$ is fitted to the experiment.

We require that the new modes ${\mit\Phi}_k$ do not destroy the correlation
energy due to the r.p.o.; this determines the extent of the
c.p.o.\  The field ${\mit\Phi} (\bmf{r})$ contains degrees of freedom which
will be thermally excited. Let $k_{\rm c}(t)$ be the average amount of the
momenta of this field
\begin{equation}
\label{e49}
k_{\rm c}(t) = \big\langle\, \big\vert \hh \bmf{\nabla} {\mit\Phi}
(\bmf{r})\hh \big\vert \, \big\rangle\,.
\end{equation}
Because of $|\hh\varphi\hh |^2 = |\hh \bar\varphi \hh |^2$ the fields
${\mit\Phi}$ and ${\mit\Phi}_{\bmfi{k}}$ do not directly influence the
spatial correlations.  However, for ${\mit\Phi}_{\bmfi{k}}\equiv 0$ the
noncondensed states $\bmf{k}$ with $\varepsilon_k<\hbar^2k_{\rm
c}^{\,2}/2\hh m$ would have a lower energy than the condensate state.
This would invalidate the assumption about the occupation pattern and
destroy the correlation energy. On the other hand, for $k\gg k_{\rm c}$
the ${\mit\Phi}_{\bmfi{k}}$ will have only minor influence on the
occupation pattern. Simplifying and quantifying this qualitative
argument leads to the following condition for c.p.o.:
\begin{equation}
\label{e50}
{\mit\Phi}_k (\bmf{r}) =\left\{ \begin{array}{ccc} 
{\mit\Phi} (\bmf{r}) & & (k<k_{\rm c})\,,\\[1mm]
 \mbox{arbitrary} && (k>k_{\rm c})\,. \end{array}
\right. \qquad {\rm (c.p.o.)}
\end{equation}
This condition means that the localized s.p.f.\  with $k<k_{\rm c}$
adopt within their range the (macroscopically defined) phase ${\mit\Phi}$ of
the condensate. The physical reason for this c.p.o.\ is that the
correlations due to the r.p.o.\ (favoured by the free energy) must not
be destroyed; in this way the real and complex p.o.\ are connected to
each other.

\section{Superfluid density}
\label{s5}

\subsection{AIBG expression}
\label{s5.1}

Using the new s.p.f.\  (\ref{e47}) and (\ref{e48}) the states (\ref{e5})
become for $t<0$:
\begin{equation}
\label{e51}
{\mit\Psi} = F\,{\cal S} \h \big(\h \varphi_0 \,\exp \,[\,{\rm i}\h\hh 
 {\mit\Phi}(\bmf{r})] \h\big)^{n_0}
 \prod_{\bmfi{k},{\rm vol}}  \big(\h \varphi_{\bmfi{k}}\, 
\exp \, [\, {\rm i}\hh\h {\mit\Phi}_{\bmfi{k}}(\bmf{r})]\h \big)^{n_k}\,.
\end{equation}
For these states we evaluate the quantum mechanical expectation value
of the current operator. Taking into account the c.p.o.\ (\ref{e50}) we
obtain
\begin{eqnarray}
\label{e52}
\bmf{j}(\bmf{r},n_k) &=& \bigg\langle {\mit\Psi} \, \bigg\vert
\;\frac{\hbar}{2\h\hh {\rm i}} \sum_{j=1}^N 
\bmf{\nabla}_{\!j}\,\delta(\bmf{r}-\bmf{r}_j) +
{\rm c.c.} \,\bigg\vert \, {\mit\Psi} \bigg\rangle 
\\[2mm]
=&&  \hspace*{-4.3mm} \frac{\rho}{N}\h\frac{\hbar}{m} \bigg[ \Big( n_0
+ \!\!\!\!\sum_{\,\,\,\,\,\,\, k<k_{\rm c}}\hspace*{-8.4mm} \int \,\,\,
 n_k \Big) \bmf{\nabla} {\mit\Phi}
(\bmf{r}) + \!\!\!\!\sum_{\,\,\,\,\,\,\, k>k_{\rm c}}
 \hspace*{-8.4mm} \int \,\,\, n_k
 \h \bmf{\nabla} {\mit\Phi}_{\bmfi{k}} (\bmf{r})\bigg]
\nonumber .
\end{eqnarray}
All derivatives of the real functions ($F$, $\varphi_0$ and
$\varphi_{\bmfi{k}}$) are cancelled by the complex conjugate (c.c.)
term; the only surviving contributions are the phase derivatives. Each
such derivative is accompanied by a factor $\langle {\mit\Psi} |\hh
\delta (\bmf{r} - \bmf{r}_j)\hh |{\mit\Psi} \rangle =\rho
(\bmf{r})/(Nm)$.  The mass density $\rho$ is assumed to be a constant
in the following.

The contributions of the nonordered phases are averaged out
statistically. This is also the case for the thermal excitations of the
${\mit\Phi}$-field, $\langle \h \bmf{\nabla} {\mit\Phi}_{\rm th}\h
\rangle =0$. A coherent nonvanishing flow can only be obtained by an
additional nonequilibrium contribution ${\mit\Phi}_{\rm s}$ to
${\mit\Phi}$,
\begin{equation}
\label{e53}
{\mit\Phi} = {\mit\Phi}_{\rm th} + {\mit\Phi}_{\rm s}\,.
\end{equation}
The flux due to ${\mit\Phi}_{\rm s}$ might be relatively stable if it is small
enough\cite{la}. This implies $|\bmf{\nabla} {\mit\Phi}_{\rm s}|\ll k_{\rm c}$ where
$k_{\rm c} = \langle\, |\bmf{\nabla} {\mit\Phi}_{\rm th} |\, \rangle$.

\begin{figure}[h]
\begin{center}
\epsfxsize=8.5cm
%\vspace{4cm}
\epsfbox{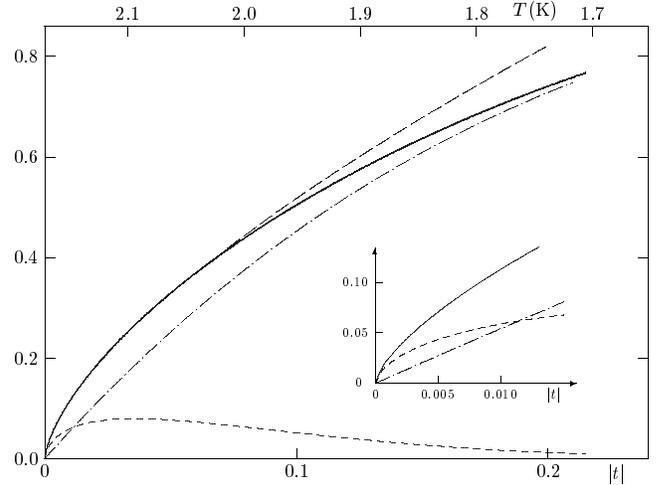}
\end{center}
\caption[]{\label{fig1}Composition of the superfluid density $\rho_{\rm
s}/\rho$ (solid line) according to Eq.\ (\ref{e54}). The ratio
$\rho_{\rm s}/\rho$ is the sum of the model condensate $\rho_0/\rho$
(short dashes) and the coherently comoving density $\rho_{\rm c}/\rho$
(dash-dotted line). Also shown is the simplest  1-parameter fit
$\rho_{\rm s}/\rho = a_1\h |t|^{2/3}$ (long dashes). In the given scale
the fitted $\rho_{\rm s}/\rho$ (solid line) coincides with the data.}
\end{figure}

Using Eq.\ (\ref{e53}) and $\langle \h\bmf{\nabla}{\mit\Phi}_{\rm th}\h\rangle =0$
we evaluate the statistical expectation value of Eq.\ (\ref{e52}).
Writing $\langle \,\bmf{j}_{\rm s}\rangle =\rho_{\rm s}\bmf{v}_{\rm s}$
with $\bmf{v}_{\rm s}=(\hbar /m)\bmf{\nabla}
{\mit\Phi}_{\rm s}$ we obtain for $\rho_{\rm s}$:
\begin{equation}
\label{e54}
\frac{\rho_{\rm s}}{\rho} = 
\frac{1}{N} \bigg( \langle n_0\rangle +
\!\!\!\!\sum_{\,\,\,\,\,\,\, k<k_{\rm c}}\hspace*{-8.4mm} \int \,\,\,\,
  \langle n_k\rangle \bigg) 
= \frac{\rho_0+\rho_{\rm c}}{\rho}\,.
\end{equation}
This is the AIBG expression for the superfluid density. In a number of
points we summarize its implications:
\begin{enumerate}
\item[1.] The density $\rho_{\rm s}$ is composed by all atoms adopting
the phase ${\mit\Phi} (\bmf{r})$. It consists of the condensate density
$\rho_0$ and the comoving density $\rho_{\rm c}$. The temperature
dependence of this composition (Fig.\ \ref{fig1}) is obtained by
fitting Eq.\ (\ref{e54}) to the data.  The comoving density $\rho_{\rm
c}$ has an internal structure which implies a nonvanishing entropy
$S_{\rm s}\ne 0$ of the super\-fluid component.
\end{enumerate}

\begin{enumerate}
\item[2.] Expression (\ref{e54}) is the simplest possibility to account
in an IBG-like model with $\langle n_0\rangle \sim |t|$ for $\rho_{\rm
s}\sim |t|^{2/3}$. Fitting Eq.\ (\ref{e54}) to the experimental
$\rho_{\rm s}$ fixes $k_{\rm c}(t)$, and consequently the difference
between the densities $\rho_0$ and $\rho_{\rm s}$ and the model
prediction for $S_{\rm s}\ne 0$.

\item[3.] The reproduction of the data with Eq.\ (\ref{e54}) requires
asymptotically $k_{\rm c}(t)\sim |t|^{2/3}$. Theoretically this
asymptotic form is made plausible in Sec.\ \ref{s4.2}. It leads to a
certain functional form of $\rho_{\rm s}$ which yields excellent fits
to the experimental data (Sec.\ \ref{s5.3}).

\item[4.] The result (\ref{e54}) contains a specific picture for the
relation between the condensate fraction $n_{\rm c}$ and the superfluid
fraction $\rho_{\rm s}/\rho$. The explanation of superfluidity is based
on the assumption of a macroscopic wave function. This connection
implies, however, an open question: What is the quantitative relation
between $n_{\rm c}(T)$ and $\rho_{\rm s}(T)/\rho$, and in particular
the relation between the $T=0$ values $n_{\rm c} \approx 0.1$ (see
Ref.\ \onlinecite{se1}) and $\rho_{\rm s}/\rho =1$\,? The
AIBG proposes the following picture: The model condensate $n_{\rm c,
M}=\langle n_0\rangle /N=\rho_0/\rho$ is the fundamental quantity. It
is depleted by the Jastrow factor $F$ in Eq.\ (\ref{e51}) to the real
condensate $n_{\rm c}$, for $T=0$ the value $n_{\rm c,M} = 1$ is reduced
to $n_{\rm c}\approx 0.1$. The factor $F$ in Eq.\ (\ref{e51}) does,
however, not deplete the current due to phase factors.  Therefore
$n_{\rm c,M}\to 1$ implies $\rho_{\rm s}/\rho\to 1$. On the other hand,
for $|t|\ll 1$ the contribution $\rho_{\rm c}\gg \rho_0$ dominates
$\rho_{\rm s}$. 
\end{enumerate}

\subsection{Effective Ginzburg-Landau model}
\label{s5.2}

Ginzburg and Sobyanin\cite{gi} have proposed an effective
Ginzburg-Landau functional for the free energy $F_{\rm GL}$ as a
function of the order parameter field $\psi$. In this ansatz singular
coefficients (like $|t|^{4/3}$ for the $|\psi|^2$ term) are introduced
in order to yield the right critical exponents. We follow this kind of
approach for investigating the relation between the critical exponents
of the order parameter and the superfluid density.  This detour is,
however, not required for the final result for $S_{\rm s}$ (see point 2
of Sec.\ \ref{s5.1}).
 
For discussing the fluctuations of the ${\mit\Phi}$-field we define the order
parameter field by
\begin{equation}
\label{e55}
\psi =\sqrt{\frac{n_0}{V}\,}\, \exp \,[\, {\rm i}\hh\h
 {\mit\Phi} (\bmf{r}) \hh  ]\,.
\end{equation}
For a definite phase the state (\ref{e51}) must be replaced by the
appropriate coherent state\cite{an}. This means that
$n_0$ in Eq.\ (\ref{e55}) has to be understood as a quantum mechanical
expectation value of the occupation number $n_0$ in such a coherent
state.

The statistical expectation value $\langle n_0\rangle \sim |t|$ can
be obtained by minimizing the common Landau energy $F_L/V=
R\,t\,|\psi|^2 +U\,|\psi|^4$ (with regular coefficients $R$ and $U$).
Adding a naive kinetic energy term $(\hbar^2/2\hh m)|\bmf{\nabla}\psi|^2$ leads
to a violation of scaling invariance and to wrong critical exponents.
The decisive feature of the AIBG expression (\ref{e54}) is that the mass
density comoving with $\bmf{\nabla} {\mit\Phi}$ is $\rho_0+\rho_{\rm c}$ instead of
$\rho_0 = m\h \langle\,|\psi|^2\rangle$ alone. This suggests the
following effective Ginzburg-Landau ansatz
\begin{equation}
\label{e56}
\frac{F_{\rm GL}}{V} = \frac{\hbar^2}{2\h m}\,\frac{\rho_{\rm
s}}{\rho_0}\, \big|\hh \bmf{\nabla}\psi\hh\big|^2 + 
R\,t\,\big|\hh\psi\hh\big|^2 + U\,\big|\hh\psi\hh\big|^4\,.
\end{equation}
Leaving away unessential constants, the asymptotic behaviour of the
mass density $\rho_{\rm c}$ in Eq.\ (\ref{e54}) is
\begin{equation}
\label{e57}
\rho_{\rm c}\sim \int_{k<k_{\rm c}} \;\frac{d^3k}{k^2+t^2} \sim k_{\rm c}\,.
\end{equation}
This, $\rho\simeq \rho_{\rm c}$ for $|t|\to 0$ and Eq.\ (\ref{e49})
determine the asymptotic kinetic energy in Eq.\ (\ref{e56}),
\begin{equation}
\label{e58}
\frac{\rho_{\rm s}}{\rho_0}\,\big\langle \,|\bmf{\nabla}\psi|^2 \big\rangle 
 =\frac{\rho_{\rm c}}{m}\,\big\langle \,|\bmf{\nabla}{\mit\Phi}|^2\hh \big\rangle 
\sim k_{\rm
c}^{\,3}\,.
\end{equation}
Scaling this kinetic part of $F_{\rm GL}$ with $\langle F_L\rangle \sim t^2$
yields
\begin{equation}
\label{e59}
k_{\rm c}\sim |t|^{2/3}\,.
\end{equation}
This implies $\rho_{\rm s}\sim\rho_{\rm c}\sim |t|^{2/3}$ and a
singular mass coefficient $\rho_{\rm s}/\rho_0\sim |t|^{-1/3}$ in Eq.\
(\ref{e56}). Therefore, Eq.\ (\ref{e56}) is an effective
Ginzburg-Landau ansatz in the same sense as the one by proposed by
Ginzburg and Sobyanin\cite{gi}. Equation (\ref{e56}) demonstrates how
the critical exponent $\beta =1/2$ (from $\langle n_0\rangle \propto
\langle |\psi|^2\rangle \sim |t|^{2\beta}$) can be connected to $\nu
=1/3$ (from $\rho_{\rm s}\sim |t|^{2\nu}$) on account of the comoving
density $\rho_{\rm c}$. The divergent mass coefficient damps the
critical fluctuations such that Eq.\ (\ref{e56}) becomes scaling
invariant (this is discussed in more detail in Ref.\ \onlinecite{f2}).
The scaling invariance implies that Eq.\ (\ref{e56}) might be used down
to $|t|=0$ and that the critical exponent of $k_{\rm c}$ and $\rho_{\rm
s}$ might be indeed exactly $2/3$.  This possibility is supported by
the excellent fits obtained from Eq.\ (\ref{e54}) with Eq.\ (\ref{e59}).

\subsection{Fit to the experimental data}
\label{s5.3}

\begin{figure}[h]
\begin{center}
\epsfxsize=8.5cm
%\vspace{4cm}
\epsfbox{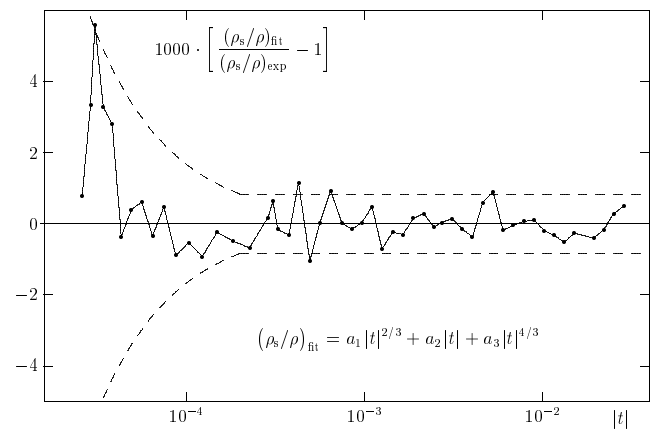}
\end{center}
\caption[]{\label{fig2}Asymptotic model fit of the superfluid density.
The differences between the 3-parameter fit formula Eq.\ (\ref{e60}) and
the data points are compared to two standard deviations (dashed
lines).}
\end{figure}

We insert the leading $\tau = a'\hh |t|$ and $k_{\rm c} = k_1\hh
|t|^{2/3}$ in Eq.\ (\ref{e54}). This yields
\begin{equation}
\label{e60}
\frac{\rho_{\rm s}}{\rho} = \h a_1\h |t|^{2/3} + a_2\h |t| + a_3\h
|t|^{4/3} +\ldots \qquad \mbox{(MAS)}\,.
\end{equation}
With 3 parameters we take this as our asymptotic model fit, called MAS.
Figure \ref{fig2} shows that MAS yields an excellent reproduction of
the data by Greywall and Ahlers\cite{a73} (GA), for saturated vapour
pressure.

The 3-parameter standard fit (SF) used by GA is
\begin{equation}
\label{e61}
\frac{\rho_{\rm s}}{\rho} = \h a_1\h |t|^{a_2} \left(1 + a_3\h
|t|^{1/2}  +\ldots \right) \qquad \mbox{(SF)}\,.
\end{equation}
We compare both fits, MAS and SF, by calculating the sums $\chi^2$ of
the quadratic deviations:
\begin{equation}
\label{e62}
\frac{\chi^2_{\rm SF}}{\chi^2_{\rm MAS}}\simeq 10 \qquad (|t|\le 0.03)\,.
\end{equation}
This shows that the reproduction shown in Fig.\ \ref{fig2} is not a
trivial result; in the considered range ($|t|\le 0.03$) the data cannot
be reproduced by SF. We remark that SF (but probably also other
3-parameter fits) fit the data in the considerably smaller range
$|t|\le 0.004$; this range is used by GA for the fit. The range $|t|\le
0.03$ seems to be appropriate for a 3-parameter fit because already a
1-parameter fit ($a_1|t|^{2/3}$) roughly reproduces the data in this
range (see Fig.\ \ref{fig1}).

The applicability range of Eq.\ (\ref{e54}) can be considerably
extended by using the following 4-parameter expansion for $\tau$ and
$k_{\rm c}$:
\begin{equation}
\label{e63}
\tau (t) =a'\,|t|,\qquad 
x_{\rm c}(t) = x_1\h |t|^{2/3} + x_2\h |t| + x_3\h |t|^{4/3}\,.
\end{equation}
The form of $x_{\rm c}=\lambda \h\hh k_{\rm c}(t)/\sqrt{4\hh\pi}$
suggests itself because it leaves the expansion (\ref{e60}) unchanged.
The 4-parameter model fit, Eq.\ (\ref{e54}) with Eq.\ (\ref{e63}),
reproduces the data down to $1\,{\rm K}$ within the experimental
errors. The fit yields the parameter values $a'=3.019$, $x_1=2.7028$,
$x_2=-0.837$ and $x_3=-3.842$; these values have been used for Fig.\
\ref{fig1}.  More details of this fit and a discussion  with respect to
the quasi-particle picture are given in Ref.\ \onlinecite{f2}.

The fit to the experimental data fixes the decomposition of $\rho_{\rm s}$
into $\rho_0$ and $\rho_{\rm c}$. This decomposition is the starting point
for the evaluation of the superfluid entropy $S_{\rm s}$.

\section{Superfluid entropy}
\label{s6}

\subsection{Model prediction}
\label{s6.1}

In $\rho_{\rm s}=\rho_0+\rho_{\rm c}$ only the condensate part $\rho_0$
corresponds to a macroscopic wave function and has thus zero entropy
content. The comoving part $\rho_{\rm c}$ is made up by different
s.p.f.\  (however with the same phase factor in Eq.\ (\ref{e48})) and
has therefore a nonvanishing entropy content. We determine this
entropy.

We start with the well-known entropy expression $S(n_k)$ for a Bose gas
with occupation numbers $n_k$. The IBG equilibrium entropy $S_{\rm
IBG}$ is obtained from this expression by $S_{\rm IBG} = \langle\h
S(n_k)\h \rangle = S(\langle n_k\rangle_{\rm IBG})$. In Sec.\
\ref{s4.1} we have discussed that for r.p.o.\ in one direction only
every second s.p.\  state is occupied. Since the total particle number
$N=\sum n_k$ is fixed we have to put twice as many atoms in every
second state (as compared to the plain IBG expression). This leads to
the following model entropy $S_{\rm M}$,
\begin{eqnarray}
\label{e64}
S_{\rm M}(T,V,N) &=&
 \frac{\kB }{2}\,\sum_{\bmfi{k}} \Big[ \big( 1 + 2\h \langle
n_k\rangle \big)\,\ln (1 + 2\langle n_k\rangle )
\nonumber\\ &&
{\qquad\quad} - 2\h \langle n_k\rangle \,\ln
(2\langle n_k\rangle )\Big]\,.
\end{eqnarray}
For $t=0$ this yields $S_{\rm M}/N = 0.96\,\kB $ instead of $S_{\rm
IBG}/N= 1.28\,\kB $. The difference $\Delta S({\rm p.o.})= S_{\rm
IBG} - S_{\rm M}$ has been compared to $\Delta E({\rm p.o})$ in Sec.\
\ref{s4.1}. For $t<0$ the $\langle n_k\rangle $ with $\tau (t)$ of Eq.\
(\ref{e63}) will be used in Eq.\ (\ref{e64}). The resulting overall
behaviour of $S_{\rm M}(T)$ is then ---in contrast to $S_{\rm
IBG}(T)$--- similar to that of the experimental entropy $S(T)$.

\begin{figure}[h]
% The parameter values for this figure are slightly different from 
% that cited after Eq.(63) and % used in Fig. 1, namely $a'=2.954$, 
% $x_1=2.6314$, $x_2=-0.824$ and  $x_3=-3.943$.
\begin{center}
\epsfxsize=8.5cm
%\vspace{4cm}
\epsfbox{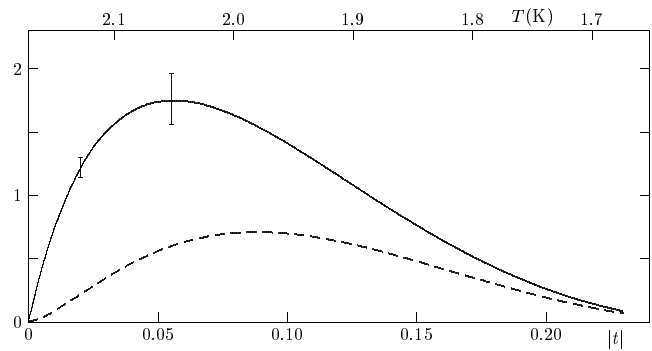}
\end{center}
\caption[]{\label{fig3}Model prediction for the superfluid entropy
$S_{\rm s}$ as a function of the temperature. The figure displays the
ratios $S_{\rm s}/S$ (dashed line) and $(S_{\rm s}/N_{\rm s})/(S/N)$
(solid line) both multiplied by a factor 100. The error bars indicate
the uncertainty of the fit parameters in Eq.\ (\ref{e63}). The
prediction becomes increasingly uncertain for $|t| > 0.1$ because the
calculation is based on an asymptotic expansion (\ref{e63}).}
\end{figure}

The entropy of the superfluid component is $S_{\rm c}=S(\rho_{\rm c})$.
For calculating $S_{\rm c}$ we have to restrict the sum in Eq.\
(\ref{e64}) by $k<k_{\rm c}$, and to account for the c.p.o.\
(\ref{e50}). Prescribing the phase field ${\mit\Phi}_{\bmfi{k}}$ for each
s.p.f.\  in Eq.\ (\ref{e48}) implies a two to one restriction for each
of the three directions of $\bmf{\bmf{\nabla}}{\mit\Phi}_{\bmfi{k}}$.
For each s.p.\ state $\bmf{k}$ this reduces the entropy by $\h\kB \ln
8\h$ (as long as $n_k\gg 1$). With these specifications we obtain
\begin{eqnarray}
\label{e65}
S_{\rm c}(T,V,N) &=& 
\frac{\kB }{2} \sum_{k< k_{\rm c}} \Big[\big( 1 + 2\langle
n_k\rangle \big)\,\ln (1+2\langle n_k\rangle ) 
\nonumber\\ &&
{\qquad\quad}  - 2\h \langle n_k\rangle \,
\ln (2\langle n_k\rangle )- \ln 8\,\Big]\,.
\end{eqnarray}
The ratio $S_{\rm c}/S_{\rm M}$ is the AIBG prediction for the
superfluid entropy fraction $S_{\rm s}/S$,
\begin{equation}
\label{e66}
\frac{S_{\rm s}}{S} =\frac{S_{\rm c}}{S_{\rm M}}\,.
\end{equation}
The calculated result is shown in Fig.\ \ref{fig3}.

\subsection{Experimental detectability}
\label{s6.2}

We show that the AIBG prediction for $S_{\rm s}\ne 0$ is at the border of
present-day experimental detectability. The most prominent experiment
showing that $S_{\rm s}=0$ (or at least $S_{\rm s}\approx 0$) is the fountain
pressure (FP) measurement. For two containers of liquid helium
connected by a superleak a temperature difference $dT$ produces a
pressure difference $dP$.  Admitting $S_{\rm s}\ne 0$ this FP is given by
\begin{equation}
\label{e67}
\left(\frac{dP}{dT}\right)_{\!\rm FP} = \, \frac{S}{V}\left(
1-\frac{S_{\rm s}/N_{\rm s}}{S/N} \right) \,.
\end{equation}
For $S_{\rm s}=0$ this reduces to the well-known London relation. The
correction term is $S_{\rm s}/S$ multiplied by $N/N_{\rm s}=\rho
/\rho_{\rm s}$; the model prediction for it is shown in Fig.\
\ref{fig3}.

The most accurate FP measurements are that by Singsaas and Ahlers
\cite{a84}. These authors assume the validity of the London relation
and interpret their experiment as an entropy measurement. For this
entropy $S_{\rm FP}$,
\begin{equation}
\label{e68}
\frac{S_{\rm FP}}{V} =\left(\frac{dP}{dT}\right)_{\rm\!FP} 
\end{equation}
they find within the errors no deviation from the true (caloric)
entropy $S$. The absolute values of $S$ near $T_\lambda$ are, however,
uncertain by about 2\% (Refs.\ \onlinecite{a84,br}).

\bigskip

\small

TABLE I. Comparison between the caloric ($S$) and the
fountain pressure ($S_{\rm FP}$) entropy. The input is the experimental
$S_{\rm FP}(t)$ and $C_{p}(t)$; the quantity $\Delta S=S_\lambda -S$ is
calculated from $C_{p}$. The last column shows the resulting marginal
evidence for $S\ne S_{\rm FP}$ or, equivalently, for $S_{\rm s}\ne 0$.

\vspace*{2mm}

\hspace*{-3.5mm}\begin{tabular}{ccccc}
\hline 
1&2&3&4&5 \\ \hline \\[-2.5mm]
$|t|$                                           &
{\footnotesize $\!\!\displaystyle
\frac{\rho\,S_{\rm FP}}{J\h{\rm cm}^{-3}\h{\rm K}^{-1}} $}       & 
{\footnotesize $\displaystyle
\frac{S_{\rm FP}}{N\kB }           $}  & 
{\footnotesize $\displaystyle
\frac{S_{\rm FP}+\Delta S}{N\kB }  $ } & 
{\footnotesize $\displaystyle
\frac{100\,(S-S_{\rm FP})}{S} \!\! \!          $ }   \\[3mm]
\hline \\[-3mm]
0.0006906 & 0.2299 & 0.7571 & $\, 0.7619\pm 0.0008$ & $\,0.07\pm 0.11$ \\
0.000790~ & 0.2296 & 0.7561 & $\, 0.7615\pm 0.0008$ & $\,0.11\pm 0.11$ \\
0.001013~ & 0.2294 & 0.7555 & $\, 0.7623\pm 0.0008$ & $\,0.02\pm 0.11$ \\
0.001794~ & 0.2275 & 0.7493 & $\, 0.7607\pm 0.0009$ & $\,0.23\pm 0.11$ \\
0.003177~ & 0.2250 & 0.7411 & $\, 0.7601\pm 0.0010$ & $\,0.30\pm 0.13$ \\
0.005662~ & 0.2208 & 0.7274 & $\, 0.7592\pm 0.0011$ & $\,0.43\pm 0.14$ \\
0.01007~~ & 0.2145 & 0.7069 & $\, 0.7596\pm 0.0013$ & $\,0.39\pm 0.17$ \\
0.03338~~ & 0.1854 & 0.6116 & $\, 0.7578\pm 0.0021$ & $\,0.74\pm 0.34$ \\
0.07973~~ & 0.1407 & 0.4648 & $\, 0.7547\pm 0.0034$ & $\,1.63 \pm 0.71$ \\ 
&&&&\\[-3mm]
\hline
\end{tabular}

\normalsize

\bigskip\smallskip

The theoretical prediction of Fig.\ \ref{fig3} suggests that one should
compare the temperature dependences of $S$ and $S_{\rm FP}$ near
$T_\lambda$ rather than the absolute values. This is done in Table I
in a number of steps:
\begin{enumerate}
\item
We start with the experimental values\cite{a84} for $S_{\rm FP}$
in column 1 and 2. By using $\rho (t)$ from equation (A1) of Ref.\
\onlinecite{ma} we relate the entropy to the particle number rather
than to the density (column 3).
\item
For column 4 we calculated $\Delta S=S_\lambda -S(T)= \int
dT\,C_P/T\h $ from the experimental specific heat $C_P$ of Ref.\
\onlinecite{a71}.  If $S_{\rm FP}$ were equal to $S$ then column 4
should show the temperature independent $S_\lambda =S(T_\lambda )$.
\item 
From the temperature dependence of column 4 we deduce $S_\lambda /N =
0.7624\,\kB $ as the limit of $S_{\rm FP}+\Delta S$ for $t\to 0^{-}$.
Then
\begin{equation}
\label{e69}
\frac{S-S_{\rm FP}}{S} =\frac{ S_\lambda -[S_{\rm FP}(t)+\Delta S(t)]}
{S_\lambda -\Delta S(t)}=\frac{S_{\rm s}/N_{\rm s}}{S/N}\,,
\end{equation}
may be calculated from column 4. The last equality in Eq.\ (\ref{e69})
follows from Eqs.\ (\ref{e67}) and (\ref{e68}). Column 5 displays the
resulting experimental evidence (or nonevidence) for $S_{\rm FP}\ne S$.
\item
The errors included in the Table are the statistical errors\cite{a84}
of 0.1\% for $S_{\rm FP}(t)$, and a 1\% error for $C_P$. The
systematic errors\cite{a84} of $S_{\rm FP}(t)$ are not included; they
should be less important because we consider only the $t$-dependence
(and not the absolute values) in a relatively small interval.
\end{enumerate}

\begin{figure}[h]
\begin{center}
\epsfxsize=8.5cm
%\vspace{4cm}
\epsfbox{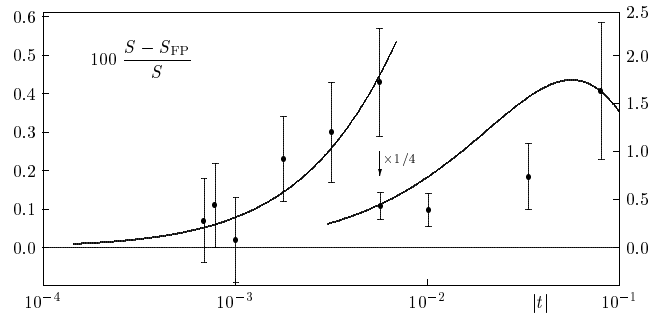}
\end{center}
\caption[]{\label{fig4}Difference between the caloric ($S$) and the
fountain pressure ($S_{\rm FP}$) entropy (dots with error bars, from
column 5 of Table 1).  This difference is compared to the model
prediction (solid line) for $(S_{\rm s}/N_{\rm s})/(S/N)$.}
\end{figure}

Figure \ref{fig4} compares column 5 of the Table with our theoretical
prediction. There seems to be some indication of a deviation $S \ne
S_{\rm FP}(t)$. The error bars (from the Table) show, however, that the
experimental evidence for such a deviation is at most marginally
significant. In any case, the compilation of Table I and Fig.\
\ref{fig4} shows that, and in which way, our theoretical prediction is
within the reach of experimental detectability.

\section{Concluding remarks}
\label{s7}

We have proposed a microscopic model for the $\lambda$-tran\-si\-tion of
liquid helium. The model uses phenomenological assumptions, in
particular the transition itself is introduced by the analogy to the
IBG. The basic idea of the AIBG is that phase ordering leads to an
extra energy $\sim \sum \nu_q^{\,2}$ which ---evaluated with the IBG
occupation numbers--- yields a logarithmic singularity. In this
model the superfluid density is not identical to the square of the
order parameter field ($\rho_{\rm s}\ne m\h |\psi|^2$). A fit to the
experimental $\rho_{\rm s}$ leads to the model prediction for a
nonvanishing superfluid entropy $S_{\rm s}$.

The model is hardly related to standard models of liquid helium which
are either microscopic variational approaches (like Refs.\
\onlinecite{fe,se2}) or the Landau-Wilson renormalization group
theory\cite{do}. Because of its novelty and originality our model
cannot compete in quality of foundation, exactness and completeness
with those other theories. It leads, however, to specific predictions
about the structure of $\rho_{\rm s}$.  These predictions are without
competition by other approaches and they can and should be tested
experimentally.  Therefore the presentation of the model seems
appropriate in spite of a number of unsolved questions concerning the
model assumptions.

In contrast to other approaches the AIBG offers also a possible
solution of the so-called microscopic problem of liquid helium. This
problem was formulated by Uhlenbeck\cite{uh}: If $\rho_{\rm s}$ is
identified with a single quantum state ($\rho_{\rm s} = m\h |\psi |^2$)
then the approach to equilibrium ($\rho_{\rm s} = \rho_{\rm s}(T)$)
cannot be understood. In the AIBG this problem is solved by the
contribution of noncondensed particles to the superfluid density.

\end {document}